\documentclass[reprint, aps, nofootinbib, prl, floatfix, superscriptaddress]{revtex4-2}

\usepackage{amsmath}
\usepackage{amssymb}
\usepackage{physics}
\usepackage{xfrac}
\usepackage{color}

\usepackage{csquotes}
\MakeOuterQuote{"}

\usepackage{hyperref}
\usepackage{graphicx}
\setkeys{Gin}{draft=false} 
\graphicspath{{figures/}}

\renewcommand\d{\ensuremath{\partial}}
\newcommand\scri{\mathcal{I}}
\newcommand\volelement{\sqrt{\abs{g}}}

\usepackage{ifdraft}
\ifdraft{
    \usepackage{setspace}
    \doublespacing
}{}

\begin{document}
\title{No Drama in 2D Black Hole Evaporation}
\author{Jonathan Barenboim}
\email{jonathan\_barenboim@sfu.ca}
\affiliation{Department of Physics, Simon Fraser University, Burnaby, BC, V5A 1S6, Canada}
\author{Andrei V.\ Frolov}
\email{frolov@sfu.ca}
\affiliation{Department of Physics, Simon Fraser University, Burnaby, BC, V5A 1S6, Canada}
\author{Gabor Kunstatter}
\email{g.kunstatter@uwinnipeg.ca}
\affiliation{Physics Department, University of Winnipeg, Winnipeg, Manitoba, R3B 2E9, Canada}
\affiliation{Department of Physics, Simon Fraser University, Burnaby, BC, V5A 1S6, Canada}

\begin{abstract}   
    We numerically calculate the spacetime describing the formation and evaporation of a regular black hole in 2D dilaton gravity. The apparent horizons evaporate smoothly in finite time to form a compact trapped region. We nevertheless see rich dynamics; an anti-trapped region forms alongside the black hole, and additional compact trapped and anti-trapped regions are formed by backreaction effects as the mass radiates away. The spacetime is asymptotically flat at future null infinity and is free of singularities and Cauchy horizons.  These results suggest that the evaporation of regular 2D black holes is unitary.
\end{abstract}

\maketitle
Little is known about the  end state of black hole  evaporation, and even the unitarity of the process is still a subject of debate. A complete rigorous analysis of this issue requires knowledge of the physics at Planck scales, but as yet there is no viable theory of quantum gravity. Hawking's derivation of black hole radiation was based on quantum field theory on a fixed classical  background. One can in principle go beyond this approximation by studying the backreaction of the Hawking radiation on the spacetime geometry, but this too proves difficult in full four-dimensional general relativity (GR).

Since gaining popularity through the model introduced by Callan, Giddings, Harvey, and Strominger (CGHS) \cite{CallanEtAlEvanescentBlack1992}, two dimensional (2D) dilaton gravity has provided toy models that offer insight into black hole dynamics. 
 For reviews see \cite{GrumillerEtAlDilatonGravity2002,MertensTuriaciSolvableModels2023} and references therein. These theories provide a more tractable alternative to GR, while sharing many key features such as black holes and Hawking radiation. Of particular interest are non-singular black hole solutions, since it is anticipated that quantum gravity should resolve the singularities that are pervasive in GR. 
 However, non-singular black hole solutions typically contain a Cauchy horizon which introduces other pathologies \cite{PoissonIsraelInternalStructure1990}.
 In the absence of a full theory of quantum gravity, 2D dilaton gravity therefore serves as a useful playground for studying quantum effects in black hole dynamics such as Hawking radiation and singularity resolution, as well as the causal structure of evaporating regular black hole spacetimes. 

In this letter we present a numerical study of the formation and evaporation of a non-singular 2D black hole based on the metric introduced by Bardeen \cite{BardeenNonsingularGeneral1968} and summarize the most interesting results about the structure of the resulting spacetime. An upcoming paper will present a more extensive analysis of the model as well as details of the numerical methods.

Our model begins with a generic 2D dilaton gravity action \cite{GrumillerEtAlDilatonGravity2002}
\begin{equation} 
    S = \frac{1}{G} \int \volelement \left[ \Phi(r) R + \Phi^{\prime\prime}(r) (\nabla r)^2 + \Phi^{\prime\prime}(r) \right] d^2 x, 
\end{equation}
where $\Phi(r)$ is a function of a dilaton field $r$, $R$ is the Ricci scalar, $G$ is the two-dimensional gravitational constant, and the prime denotes differentiation with respect to $r$. This form for the action is chosen so that the metric takes an asymptotically flat, Schwarzschild-like form, but can be related to an action with arbitrary kinetic and potential terms through a Weyl transformation and/or field redefinition \cite{MartinezEtAlExactDirac1994, GrumillerEtAlDilatonGravity2002}. 

There is a unique vacuum solution up to a parameter $M$, the ADM mass \cite{Louis-MartinezKunstatterBirkhoffTheorem1994},
\begin{equation} \label{eq:vac_metric_tr_coords}
    ds^2 = -\left(1 - \frac{2M}{J(r)}\right) dt^2 + \left(1 - \frac{2M}{J(r)}\right)^{-1} dr^2,
\end{equation}
where $J(r) = \Phi^\prime(r)$.

The structure of the vacuum spacetime is specified by the metric function $J(r)$. To construct a non-singular black hole we require an analytic function $J(r)$ with the following properties:

(i) The metric reduces to the Schwarzschild metric in spherically symmetric gravity (SSG) sufficiently far from the center, i.e. $J(r) \sim r$ as $r \rightarrow \infty$; 

(ii) The curvature is finite everywhere, most easily ensured by requiring that $J > 0$ and that it diverge at least as fast as $r^{-2}$ as $r \rightarrow 0$;

(iii) The center of the black hole is replaced by a de~Sitter core with the curvature approaching a constant finite value near $r = 0$, which requires $J \sim r^{-2}$ as $r \rightarrow 0$;

(iv) There are at most two horizons for any mass, which requires that $J(r)$ have a single local minimum. 

As a specific example we focus on the Bardeen metric given by 
\begin{equation}\label{eq:bardeen_metric_function}
    J = \frac{(r^2 + l^2)^{\sfrac{3}{2}}}{r^2},
\end{equation}
where $l$ is a length parameter that determines the scale at which the spacetime transitions from Schwarzschild to de~Sitter, typically taken to be around the Planck scale. Without loss of generality we set $l = 1$. In analogy to SSG, where the dilaton field is directly related to the areal radius, we interpret $r$ as a radial coordinate and accordingly limit the solution to the region $r \ge 0$.

To model a black hole formed by the collapse of an infinitely thin shell two different vacuum metrics, an interior solution with $M=0$ and an exterior solution with $M>0$, are joined along the null shell trajectory $v = v_0$, where we use conformal null coordinates
\begin{equation}
    ds^2 = -e^{2\rho} \, du \, dv. 
    \label{eq:DoubleNull}
\end{equation}
The resulting stress-energy tensor is a shock wave,
\begin{equation}
    T_{uu} = 0, \qquad T_{vv} = M \delta (v - v_0), 
\end{equation}
with $T_{uv}$ identically 0 in the classical theory.

Hawking radiation is modelled by adding the stress-energy tensor corresponding to the one-loop conformal (trace) anomaly in 2D \cite{ChristensenFullingTraceAnomalies1977}, 
\begin{equation}
    T \equiv g^{\mu\nu}T_{\mu\nu} = \mu R,
    \label{eq:ConformalAnomaly}
\end{equation}
where $\mu = N \hbar / 24$ is a parameter characterizing the strength of the quantum effects with $N$ scalar fields\footnote{In the large $N$ limit the expression (\ref{eq:ConformalAnomaly}) for the conformal anomaly is exact.}. Henceforth we work in units where $G = \hbar = 1$.

The dynamic equations of motion for the metric and dilaton can be written as 
\begin{subequations}\label{eq:eoms_form2}
    \begin{align}\label{eq:eoms_form2_a}
        J Q \, \nabla^2 r - 2 P \mathcal{M} = 0, \\
        J Q R + 2 \Pi \mathcal{M} = 0
    \end{align}
\end{subequations}
where
\begin{equation}
    \begin{gathered}
    Q = J - 2\mu \frac{J^\prime}{J},\quad P = J^\prime - \mu \frac{J^{\prime\prime}}{J},\quad \Pi = J^{\prime\prime} - 2\frac{(J^\prime)^2}{J},
    \end{gathered}
\end{equation}
and
\begin{equation}\label{eq:mass_function}
    \mathcal{M} = \frac{J}{2} \left[1 - (\nabla r)^2\right]
\end{equation}
is a generalized Misner-Sharp mass function.

In double null coordinates (\ref{eq:DoubleNull}) the semi-classical dynamic equations of motion become
\begin{subequations}\label{eq:eom_null_coords}
\begin{align}
    J\d_u \d_v r + J^\prime \d_u r \, \d_v r + \frac{1}{4} e^{2\rho} J^\prime = -2\mu \d_u \d_v \rho, \\ 
    2 J \d_u \d_v \rho + J^{\prime\prime} \d_u r \, \d_v r + 2 J^\prime \d_u \d_v r + \frac{1}{4} e^{2\rho} J^{\prime\prime} = 0,
\end{align}
\end{subequations}
which are solved after transforming to compactified coordinates. The computational grid cannot extend all the way to future/past null infinity ($\scri^\pm$), but is chosen to cover all of the interesting dynamics. The remaining two equations,
\begin{subequations}
\begin{align}
    & 2 J \d_u \rho \d_u r - J \d_u \d_u r = 2 \mu (\d_u \d_u \rho - \d_u \rho \d_u \rho + t_u),\\
    & 2 J \d_v \rho \d_v r - J \d_v \d_v r = 2 \mu (\d_v \d_v \rho - \d_v \rho \d_v \rho + t_v),
\end{align}
\end{subequations}
are constraints that determine the boundary conditions, which are taken as the classical solution along the shell and on the initial $u$ slice, chosen sufficiently far from the black hole so that the classical solution does not give rise to any radiation at past null infinity. In the coordinates used these conditions fix the functions of integration, $t_u(u)$ and $t_v(v)$, to be 0.

\begin{figure}[t]
    \centering
    \includegraphics[width=\linewidth]{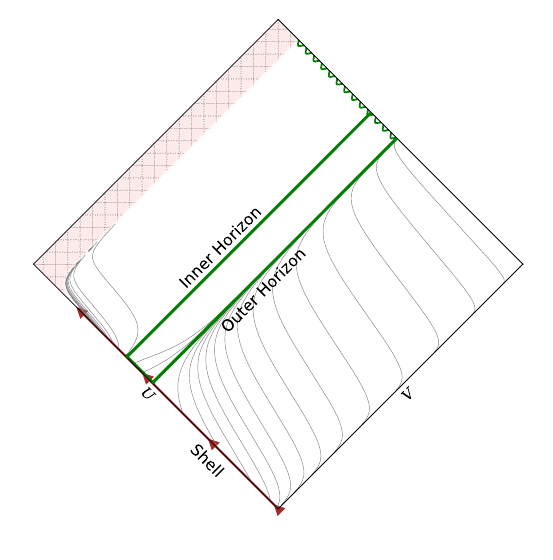}
    \caption{Conformal diagram of the classical Bardeen black hole with mass $M=1.33$. Only the solution exterior to the shell, lying along the bottom left axis, is shown. The thick green lines mark the apparent horizons and the thin grey lines are curves of constant $r$. The hatched light pink region at the upper left edge of the plot is $r < 0$ and is not part of the spacetime. The rippled line at the top shows where the spacetime connects to the next region through the Cauchy horizon at $v = \infty$.}
    \label{fig:bardeen_vac}
\end{figure} 

The classical (non-radiating) spacetime is similar to the Reissner-Nordstrom spacetime, in that it features two null horizons bounding a trapped region that ends on a Cauchy horizon, as shown in Fig.~\ref{fig:bardeen_vac}, beyond which lies an additional untrapped region. Thus our coordinates do not cover the complete classical spacetime, but can be analytically continued past the Cauchy horizon.

When radiation is added, the outer horizon shrinks while the inner horizon grows. The trapped region between them contracts, closing off smoothly in finite time. The radius and mass function at the point the horizons meet is independent of the initial mass of the collapsing shell, and depends only on the parameter $\mu$. This can be seen from Eqs.~(\ref{eq:eoms_form2}--\ref{eq:mass_function}) by noting that $\nabla^2 r = 0 = (\nabla r)^2$ when the horizons meet. The former condition requires $P(r)=0$, which implicitly determines $r(\mu)$.

A singularity occurs whenever $Q(r)=0$, where the order of the PDEs changes, and generally corresponds to a curvature singularity. There is always a positive solution to $Q(r)=0$ for singular models, but remarkably this singularity is avoided for regularized models when $\mu$ is below a critical value (which is always positive when $J(r)$ satisfies the conditions outlined above) because $Q(r)=0$ has no real, positive solutions. However, if the radiation strength $\mu$ is large compared to the regularization scale $l$ ($\mu_{\text{crit}} \approx 6.55\, l^2$ for the Bardeen model), curvature singularities appear in the radiating solutions even when there is no singularity classically. We thus consider the singularity as a breakdown of the semi-classical approximation.

\begin{figure}[t]
    \centering
    \includegraphics[width=\linewidth]{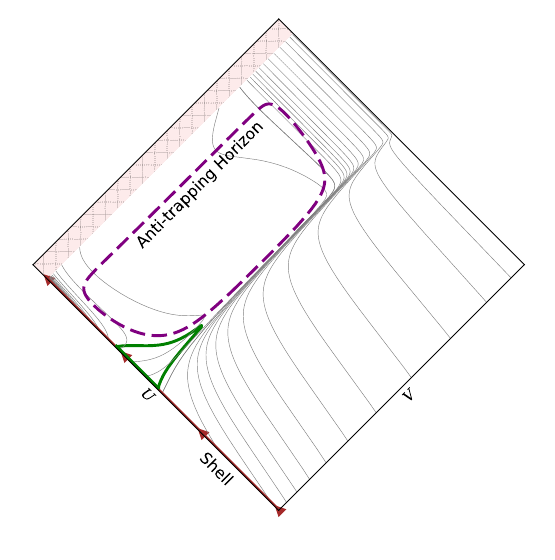}
    \caption{Evaporating Bardeen black hole with $M=1.33,~\mu=0.5$. The anti-trapping horizon $\d_u r = 0$ is shown by the dashed purple curve.}
    \label{fig:bardeen}
\end{figure}

In addition to the trapped region an anti-trapped region, where both null expansions are positive, also forms when the shell carries sufficient mass. The critical mass for white hole formation is similar to that of the black hole, approaching the classical value $M_{\text{crit, classical}} = \sqrt{27/16}$ as $\mu \rightarrow 0$ and decreasing with larger radiation strength, becoming 0 at the critical value for $\mu$. When the strength of the radiation is increased, the backreaction effects cause a series of trapped and anti-trapped regions to form after the initial black hole evaporates.

The anti-trapped region appears to be generic in non-singular models. We have seen similar structure in simulations of other non-singular metrics, and evidence of an anti-trapped region was seen in previous work on other regular dilaton models, though not identified as such.\footnote{See for example Fig.~2 in \cite{DibaLoweNearextremalBlack2002} where the  "additional contour" for $\tilde{\phi}_0$ being spacelike implies the presence of an additional horizon.}

\begin{figure}[t]
    \centering
    \includegraphics[width=\linewidth]{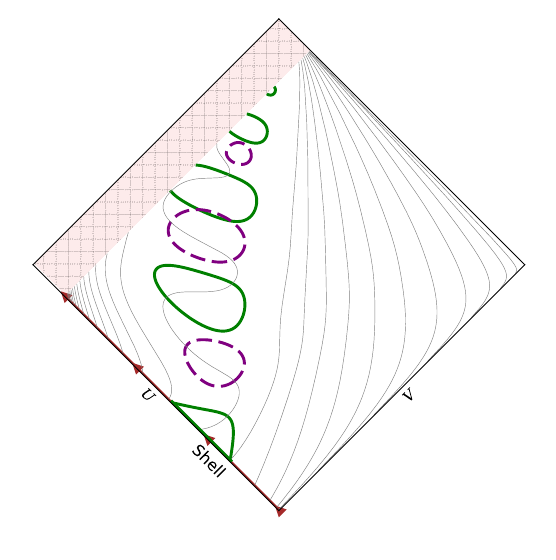}
    \caption{Bardeen black hole, $M=0.77,~\mu=5.5$. A series of trapped and anti-trapped regions form.}
    \label{fig:MultipleTrappedRegions}
\end{figure}

In the quasi-static approximation for Hawking radiation an extremal black hole does not emit any radiation. Thus it may be expected that when the inner and outer horizons meet a static extremal horizon may form. Our simulations suggest that this is not the case.  To confirm this we calculate the outgoing null expansion $\partial_v r$ on a few slices $v > v_{\text{meet}}$. A minimum of 0 would indicate the presence of a single apparent horizon and an extremal black hole. We find that while $\partial_v r$ has a small, positive minimum for all $v$, convergence tests show that this minimum does not decrease with improved resolution or numerical accuracy, confirming $\partial_v r$ does not have any zeros and there is no extremal horizon remaining once the two horizons merge. 

\begin{figure}[hbpt]
    \centering
    \includegraphics[width=\linewidth]{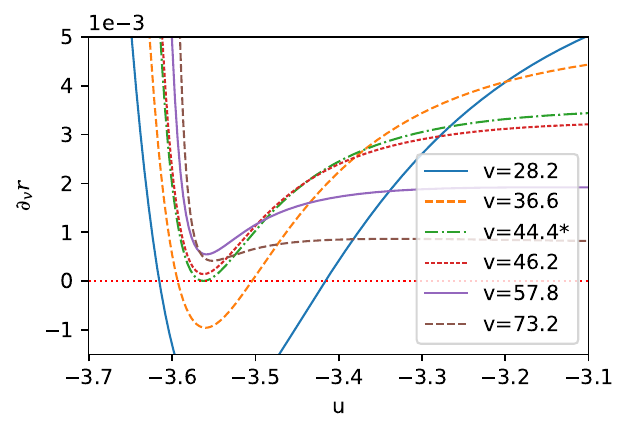}
    \caption{Horizon function $\partial_v r$ on selected slices of $v$. When a trapped region is present $\partial_v r$ crosses 0 twice, indicating the presence of two apparent horizons. The black hole disappears when the apparent horizons meet at a single point ($v$ slice marked by an asterisk). At later $v$ the horizon function remains positive, signifying that there is no apparent horizon present.}
    \label{fig:drdv_slices}
\end{figure}

Additionally, in contrast to the classical solution, the conformal factor $e^{2\rho}$ remains finite as $v \rightarrow \infty$, indicating that the semi-classical spacetime is complete in the sense that $v = \infty$ corresponds to a point on $\scri^+$ for all $u$. Our results therefore imply the absence of a Cauchy horizon, in contrast to earlier work which suggested that the endpoint of evaporation for regular black holes is an extremal configuration \cite{LoweOLoughlinNonsingularBlack1993, CadoniEtAlEvaporationInformation2023, DibaLoweNearextremalBlack2002}.  

Despite resolving the curvature singularity, the consistency of two horizon regular black hole models is potentially threatened by the prospect of mass inflation associated with Cauchy horizons \cite{PoissonIsraelInnerhorizonInstability1989}. In a simplified version of the problem, when perturbations are modelled as crossing ingoing and outgoing thin shells \cite{BarrabesEtAlCollisionLightlike1990, Carballo-RubioEtAlViabilityRegular2018} the final mass function behaves as 
\begin{equation}
    \Delta \mathcal{M} \propto \frac{-\mathcal{M}_{\text{in}}}{(\nabla r)^2}
\end{equation}
evaluated at the point the shells cross, where $\mathcal{M}_{\text{in}}$ is the mass of the ingoing shell, typically taken to follow an inverse power law. In the classical solution $(\nabla r)^2$ vanishes exponentially near the Cauchy horizon at $v = \infty$, causing the mass to diverge and nullifying the self-consistency of the model. When radiation is added there is no horizon present and $(\nabla r)^2$ remains finite at large $v$, so $\Delta \mathcal{M}$ vanishes rather than diverges. However, as pointed out in \cite{Carballo-RubioEtAlMassInflation2024}, mass inflation may pose a problem even if no Cauchy horizon is present if the evolution proceeds adiabatically as the energy density will build up before a macroscopic black hole has time to evaporate. We are unable to address this issue because numerical limitations do not allow us to study large black holes.

In summary, our numerical simulations show richer dynamics for an evaporating black hole than previously seen, with additional trapped and anti-trapped regions formed by radiation reaction. While we have focused on a specific model, the work in this paper can easily be generalized thanks to the choice of parameterization of the action. Preliminary investigations of other models show similar structure to the Bardeen model, suggesting that these features are generic in non-singular solutions. 

We see no evidence of drama in the form of singularities or Cauchy horizons, potentially ameliorating the mass inflation problem generally associated with two-horizon black holes. As we expect all apparent horizons to vanish in finite time and there is no mechanism to halt the radiation, all energy and information should eventually escape to infinity. These results support the expectation that the evaporation of regular black holes is a unitary process. 
However, whether the information can escape from the black hole within the Page time \cite{PageInformationBlack1993} remains an open question. Additionally, with this model we were only able to simulate a small range of microscopic black holes. At larger masses the simulation breaks down and we are not able to resolve the trapped region closing off. \\[10pt]

\leftline{\bf Acknowledgments}

GK is very grateful to Jonathan Ziprick and Tim Taves for helpful conversations. Authors gratefully acknowledge that this research was supported in part by Discovery Grants number 2020-05346 (AF) and 2018-0409 (GK) from the Natural Sciences and Engineering Research Council of Canada.

\bibliography{references}

\begin{thebibliography}{16}%
\makeatletter
\providecommand \@ifxundefined [1]{%
 \@ifx{#1\undefined}
}%
\providecommand \@ifnum [1]{%
 \ifnum #1\expandafter \@firstoftwo
 \else \expandafter \@secondoftwo
 \fi
}%
\providecommand \@ifx [1]{%
 \ifx #1\expandafter \@firstoftwo
 \else \expandafter \@secondoftwo
 \fi
}%
\providecommand \natexlab [1]{#1}%
\providecommand \enquote  [1]{``#1''}%
\providecommand \bibnamefont  [1]{#1}%
\providecommand \bibfnamefont [1]{#1}%
\providecommand \citenamefont [1]{#1}%
\providecommand \href@noop [0]{\@secondoftwo}%
\providecommand \href [0]{\begingroup \@sanitize@url \@href}%
\providecommand \@href[1]{\@@startlink{#1}\@@href}%
\providecommand \@@href[1]{\endgroup#1\@@endlink}%
\providecommand \@sanitize@url [0]{\catcode `\\12\catcode `\$12\catcode
  `\&12\catcode `\#12\catcode `\^12\catcode `\_12\catcode `\%12\relax}%
\providecommand \@@startlink[1]{}%
\providecommand \@@endlink[0]{}%
\providecommand \url  [0]{\begingroup\@sanitize@url \@url }%
\providecommand \@url [1]{\endgroup\@href {#1}{\urlprefix }}%
\providecommand \urlprefix  [0]{URL }%
\providecommand \Eprint [0]{\href }%
\providecommand \doibase [0]{https://doi.org/}%
\providecommand \selectlanguage [0]{\@gobble}%
\providecommand \bibinfo  [0]{\@secondoftwo}%
\providecommand \bibfield  [0]{\@secondoftwo}%
\providecommand \translation [1]{[#1]}%
\providecommand \BibitemOpen [0]{}%
\providecommand \bibitemStop [0]{}%
\providecommand \bibitemNoStop [0]{.\EOS\space}%
\providecommand \EOS [0]{\spacefactor3000\relax}%
\providecommand \BibitemShut  [1]{\csname bibitem#1\endcsname}%
\let\auto@bib@innerbib\@empty
\bibitem [{\citenamefont {Callan}\ \emph {et~al.}(1992)\citenamefont {Callan},
  \citenamefont {Giddings}, \citenamefont {Harvey},\ and\ \citenamefont
  {Strominger}}]{CallanEtAlEvanescentBlack1992}%
  \BibitemOpen
  \bibfield  {author} {\bibinfo {author} {\bibfnamefont {C.~G.}\ \bibnamefont
  {Callan}}, \bibinfo {author} {\bibfnamefont {S.~B.}\ \bibnamefont
  {Giddings}}, \bibinfo {author} {\bibfnamefont {J.~A.}\ \bibnamefont
  {Harvey}},\ and\ \bibinfo {author} {\bibfnamefont {A.}~\bibnamefont
  {Strominger}},\ }\bibfield  {title} {\bibinfo {title} {Evanescent black
  holes},\ }\href {https://doi.org/10.1103/PhysRevD.45.R1005} {\bibfield
  {journal} {\bibinfo  {journal} {Physical Review D}\ }\textbf {\bibinfo
  {volume} {45}},\ \bibinfo {pages} {R1005} (\bibinfo {year}
  {1992})}\BibitemShut {NoStop}%
\bibitem [{\citenamefont {Grumiller}\ \emph {et~al.}(2002)\citenamefont
  {Grumiller}, \citenamefont {Kummer},\ and\ \citenamefont
  {Vassilevich}}]{GrumillerEtAlDilatonGravity2002}%
  \BibitemOpen
  \bibfield  {author} {\bibinfo {author} {\bibfnamefont {D.}~\bibnamefont
  {Grumiller}}, \bibinfo {author} {\bibfnamefont {W.}~\bibnamefont {Kummer}},\
  and\ \bibinfo {author} {\bibfnamefont {D.}~\bibnamefont {Vassilevich}},\
  }\bibfield  {title} {\bibinfo {title} {Dilaton gravity in two dimensions},\
  }\href {https://doi.org/10.1016/S0370-1573(02)00267-3} {\bibfield  {journal}
  {\bibinfo  {journal} {Physics Reports}\ }\textbf {\bibinfo {volume} {369}},\
  \bibinfo {pages} {327} (\bibinfo {year} {2002})}\BibitemShut {NoStop}%
\bibitem [{\citenamefont {Mertens}\ and\ \citenamefont
  {Turiaci}(2023)}]{MertensTuriaciSolvableModels2023}%
  \BibitemOpen
  \bibfield  {author} {\bibinfo {author} {\bibfnamefont {T.~G.}\ \bibnamefont
  {Mertens}}\ and\ \bibinfo {author} {\bibfnamefont {G.~J.}\ \bibnamefont
  {Turiaci}},\ }\bibfield  {title} {\bibinfo {title} {Solvable models of
  quantum black holes: A review on {{Jackiw}}--{{Teitelboim}} gravity},\ }\href
  {https://doi.org/10.1007/s41114-023-00046-1} {\bibfield  {journal} {\bibinfo
  {journal} {Living Reviews in Relativity}\ }\textbf {\bibinfo {volume} {26}},\
  \bibinfo {pages} {4} (\bibinfo {year} {2023})}\BibitemShut {NoStop}%
\bibitem [{\citenamefont {Poisson}\ and\ \citenamefont
  {Israel}(1990)}]{PoissonIsraelInternalStructure1990}%
  \BibitemOpen
  \bibfield  {author} {\bibinfo {author} {\bibfnamefont {E.}~\bibnamefont
  {Poisson}}\ and\ \bibinfo {author} {\bibfnamefont {W.}~\bibnamefont
  {Israel}},\ }\bibfield  {title} {\bibinfo {title} {Internal structure of
  black holes},\ }\href {https://doi.org/10.1103/PhysRevD.41.1796} {\bibfield
  {journal} {\bibinfo  {journal} {Physical Review D}\ }\textbf {\bibinfo
  {volume} {41}},\ \bibinfo {pages} {1796} (\bibinfo {year}
  {1990})}\BibitemShut {NoStop}%
\bibitem [{\citenamefont {Bardeen}(1968)}]{BardeenNonsingularGeneral1968}%
  \BibitemOpen
  \bibfield  {author} {\bibinfo {author} {\bibfnamefont {J.}~\bibnamefont
  {Bardeen}},\ }\href@noop {} {\emph {\bibinfo {title} {Non-Singular General
  Relativistic Gravitational Collapse}}}\ (\bibinfo {year} {1968})\ p.~\bibinfo
  {pages} {87}\BibitemShut {NoStop}%
\bibitem [{\citenamefont {Martinez}\ \emph {et~al.}(1994)\citenamefont
  {Martinez}, \citenamefont {Gegenberg},\ and\ \citenamefont
  {Kunstatter}}]{MartinezEtAlExactDirac1994}%
  \BibitemOpen
  \bibfield  {author} {\bibinfo {author} {\bibfnamefont {D.~L.}\ \bibnamefont
  {Martinez}}, \bibinfo {author} {\bibfnamefont {J.}~\bibnamefont
  {Gegenberg}},\ and\ \bibinfo {author} {\bibfnamefont {G.}~\bibnamefont
  {Kunstatter}},\ }\bibfield  {title} {\bibinfo {title} {Exact {{Dirac
  Quantization}} of {{All}} 2-{{D Dilaton Gravity Theories}}},\ }\href
  {https://doi.org/10.1016/0370-2693(94)90463-4} {\bibfield  {journal}
  {\bibinfo  {journal} {Physics Letters B}\ }\textbf {\bibinfo {volume}
  {321}},\ \bibinfo {pages} {193} (\bibinfo {year} {1994})},\ \Eprint
  {https://arxiv.org/abs/gr-qc/9309018} {arXiv:gr-qc/9309018} \BibitemShut
  {NoStop}%
\bibitem [{\citenamefont {{Louis-Martinez}}\ and\ \citenamefont
  {Kunstatter}(1994)}]{Louis-MartinezKunstatterBirkhoffTheorem1994}%
  \BibitemOpen
  \bibfield  {author} {\bibinfo {author} {\bibfnamefont {D.}~\bibnamefont
  {{Louis-Martinez}}}\ and\ \bibinfo {author} {\bibfnamefont {G.}~\bibnamefont
  {Kunstatter}},\ }\bibfield  {title} {\bibinfo {title} {Birkhoff's theorem in
  two-dimensional dilaton gravity},\ }\href
  {https://doi.org/10.1103/PhysRevD.49.5227} {\bibfield  {journal} {\bibinfo
  {journal} {Physical Review D}\ }\textbf {\bibinfo {volume} {49}},\ \bibinfo
  {pages} {5227} (\bibinfo {year} {1994})}\BibitemShut {NoStop}%
\bibitem [{\citenamefont {Christensen}\ and\ \citenamefont
  {Fulling}(1977)}]{ChristensenFullingTraceAnomalies1977}%
  \BibitemOpen
  \bibfield  {author} {\bibinfo {author} {\bibfnamefont {S.~M.}\ \bibnamefont
  {Christensen}}\ and\ \bibinfo {author} {\bibfnamefont {S.~A.}\ \bibnamefont
  {Fulling}},\ }\bibfield  {title} {\bibinfo {title} {Trace anomalies and the
  {{Hawking}} effect},\ }\href {https://doi.org/10.1103/PhysRevD.15.2088}
  {\bibfield  {journal} {\bibinfo  {journal} {Physical Review D}\ }\textbf
  {\bibinfo {volume} {15}},\ \bibinfo {pages} {2088} (\bibinfo {year}
  {1977})}\BibitemShut {NoStop}%
\bibitem [{\citenamefont {Diba}\ and\ \citenamefont
  {Lowe}(2002)}]{DibaLoweNearextremalBlack2002}%
  \BibitemOpen
  \bibfield  {author} {\bibinfo {author} {\bibfnamefont {K.}~\bibnamefont
  {Diba}}\ and\ \bibinfo {author} {\bibfnamefont {D.~A.}\ \bibnamefont
  {Lowe}},\ }\bibfield  {title} {\bibinfo {title} {Near-extremal black hole
  evaporation in asymptotically flat spacetime},\ }\href
  {https://doi.org/10.1103/PhysRevD.66.024039} {\bibfield  {journal} {\bibinfo
  {journal} {Physical Review D}\ }\textbf {\bibinfo {volume} {66}},\ \bibinfo
  {pages} {024039} (\bibinfo {year} {2002})}\BibitemShut {NoStop}%
\bibitem [{\citenamefont {Lowe}\ and\ \citenamefont
  {O'Loughlin}(1993)}]{LoweOLoughlinNonsingularBlack1993}%
  \BibitemOpen
  \bibfield  {author} {\bibinfo {author} {\bibfnamefont {D.~A.}\ \bibnamefont
  {Lowe}}\ and\ \bibinfo {author} {\bibfnamefont {M.}~\bibnamefont
  {O'Loughlin}},\ }\bibfield  {title} {\bibinfo {title} {Nonsingular black hole
  evaporation and ``stable'' remnants},\ }\href
  {https://doi.org/10.1103/PhysRevD.48.3735} {\bibfield  {journal} {\bibinfo
  {journal} {Physical Review D}\ }\textbf {\bibinfo {volume} {48}},\ \bibinfo
  {pages} {3735} (\bibinfo {year} {1993})}\BibitemShut {NoStop}%
\bibitem [{\citenamefont {Cadoni}\ \emph {et~al.}(2023)\citenamefont {Cadoni},
  \citenamefont {Oi},\ and\ \citenamefont
  {Sanna}}]{CadoniEtAlEvaporationInformation2023}%
  \BibitemOpen
  \bibfield  {author} {\bibinfo {author} {\bibfnamefont {M.}~\bibnamefont
  {Cadoni}}, \bibinfo {author} {\bibfnamefont {M.}~\bibnamefont {Oi}},\ and\
  \bibinfo {author} {\bibfnamefont {A.~P.}\ \bibnamefont {Sanna}},\ }\bibfield
  {title} {\bibinfo {title} {Evaporation and information puzzle for {{2D}}
  nonsingular asymptotically flat black holes},\ }\href
  {https://doi.org/10.1007/JHEP06(2023)211} {\bibfield  {journal} {\bibinfo
  {journal} {Journal of High Energy Physics}\ }\textbf {\bibinfo {volume}
  {2023}},\ \bibinfo {pages} {211} (\bibinfo {year} {2023})},\ \Eprint
  {https://arxiv.org/abs/2303.05557} {arXiv:2303.05557 [hep-th]} \BibitemShut
  {NoStop}%
\bibitem [{\citenamefont {Poisson}\ and\ \citenamefont
  {Israel}(1989)}]{PoissonIsraelInnerhorizonInstability1989}%
  \BibitemOpen
  \bibfield  {author} {\bibinfo {author} {\bibfnamefont {E.}~\bibnamefont
  {Poisson}}\ and\ \bibinfo {author} {\bibfnamefont {W.}~\bibnamefont
  {Israel}},\ }\bibfield  {title} {\bibinfo {title} {Inner-horizon instability
  and mass inflation in black holes},\ }\href
  {https://doi.org/10.1103/PhysRevLett.63.1663} {\bibfield  {journal} {\bibinfo
   {journal} {Physical Review Letters}\ }\textbf {\bibinfo {volume} {63}},\
  \bibinfo {pages} {1663} (\bibinfo {year} {1989})}\BibitemShut {NoStop}%
\bibitem [{\citenamefont {Barrabes}\ \emph {et~al.}(1990)\citenamefont
  {Barrabes}, \citenamefont {Israel},\ and\ \citenamefont
  {Poisson}}]{BarrabesEtAlCollisionLightlike1990}%
  \BibitemOpen
  \bibfield  {author} {\bibinfo {author} {\bibfnamefont {C.}~\bibnamefont
  {Barrabes}}, \bibinfo {author} {\bibfnamefont {W.}~\bibnamefont {Israel}},\
  and\ \bibinfo {author} {\bibfnamefont {E.}~\bibnamefont {Poisson}},\
  }\bibfield  {title} {\bibinfo {title} {Collision of light-like shells and
  mass inflation in rotating black holes},\ }\href
  {https://doi.org/10.1088/0264-9381/7/12/002} {\bibfield  {journal} {\bibinfo
  {journal} {Classical and Quantum Gravity}\ }\textbf {\bibinfo {volume} {7}},\
  \bibinfo {pages} {L273} (\bibinfo {year} {1990})}\BibitemShut {NoStop}%
\bibitem [{\citenamefont {{Carballo-Rubio}}\ \emph {et~al.}(2018)\citenamefont
  {{Carballo-Rubio}}, \citenamefont {Di~Filippo}, \citenamefont {Liberati},
  \citenamefont {Pacilio},\ and\ \citenamefont
  {Visser}}]{Carballo-RubioEtAlViabilityRegular2018}%
  \BibitemOpen
  \bibfield  {author} {\bibinfo {author} {\bibfnamefont {R.}~\bibnamefont
  {{Carballo-Rubio}}}, \bibinfo {author} {\bibfnamefont {F.}~\bibnamefont
  {Di~Filippo}}, \bibinfo {author} {\bibfnamefont {S.}~\bibnamefont
  {Liberati}}, \bibinfo {author} {\bibfnamefont {C.}~\bibnamefont {Pacilio}},\
  and\ \bibinfo {author} {\bibfnamefont {M.}~\bibnamefont {Visser}},\
  }\bibfield  {title} {\bibinfo {title} {On the viability of regular black
  holes},\ }\href {https://doi.org/10.1007/JHEP07(2018)023} {\bibfield
  {journal} {\bibinfo  {journal} {Journal of High Energy Physics}\ }\textbf
  {\bibinfo {volume} {2018}},\ \bibinfo {pages} {23} (\bibinfo {year}
  {2018})}\BibitemShut {NoStop}%
\bibitem [{\citenamefont {{Carballo-Rubio}}\ \emph {et~al.}(2024)\citenamefont
  {{Carballo-Rubio}}, \citenamefont {Di~Filippo}, \citenamefont {Liberati},\
  and\ \citenamefont {Visser}}]{Carballo-RubioEtAlMassInflation2024}%
  \BibitemOpen
  \bibfield  {author} {\bibinfo {author} {\bibfnamefont {R.}~\bibnamefont
  {{Carballo-Rubio}}}, \bibinfo {author} {\bibfnamefont {F.}~\bibnamefont
  {Di~Filippo}}, \bibinfo {author} {\bibfnamefont {S.}~\bibnamefont
  {Liberati}},\ and\ \bibinfo {author} {\bibfnamefont {M.}~\bibnamefont
  {Visser}},\ }\href@noop {} {\bibinfo {title} {Mass inflation without
  {{Cauchy}} horizons}} (\bibinfo {year} {2024}),\ \Eprint
  {https://arxiv.org/abs/2402.14913} {arXiv:2402.14913 [gr-qc, physics:hep-th]}
  \BibitemShut {NoStop}%
\bibitem [{\citenamefont {Page}(1993)}]{PageInformationBlack1993}%
  \BibitemOpen
  \bibfield  {author} {\bibinfo {author} {\bibfnamefont {D.~N.}\ \bibnamefont
  {Page}},\ }\bibfield  {title} {\bibinfo {title} {Information in black hole
  radiation},\ }\href {https://doi.org/10.1103/PhysRevLett.71.3743} {\bibfield
  {journal} {\bibinfo  {journal} {Physical Review Letters}\ }\textbf {\bibinfo
  {volume} {71}},\ \bibinfo {pages} {3743} (\bibinfo {year}
  {1993})}\BibitemShut {NoStop}%
\end{thebibliography}%

\end{document}